\documentstyle[epsf]{article}
\begin{document}
\begin{center}
{\Large \bf Baryons and Nuclei in the Large $N_c$ Limit}
\end{center}
\vspace{0.3cm}

\begin{center}

{Dan Olof Riska}
\end{center}
\vspace{0.3cm}

{\it Department of Physics, POB 9, 00014 University of Helsinki,
Finland}

\vspace{0.5cm}

{\bf Abstract}
\vspace{0.5cm}
  
The relation between the Skyrme model and the constituent quark model,
which appears in the large $N_c$ limit is described. Examples of
similarity in the predicted phenomenology for baryons are shown.
Finally the application to nuclei is discussed.

\vspace{1.0cm}

\section{Introduction}

Judah M. Eisenberg was elegant in appearance as well as in
expression. His taste for elegance carried over into his research, and
hence it was most natural that his attention should also be drawn to
Skyrme's topological soliton model for the baryons. He published 21
papers on the application of this model, the mathematical beauty of
which trancends its phenomenological utility.\\

In the non-perturbative region, which comprises most low energy baryon
structure and all of nuclear structure, the large color limit of QCD
has proven of great utility \cite{Hooft,Jenkins}. In this limit those
Feynman diagrams, which have the largest $N_c$ factor in dominate the
$S$-matrix. Those diagrams involve only planar gluons, which then may
be replaced by $q\bar q$ pair lines, and as a consequence all surviving
diagrams admit a meson exchange interpretation. \\

In the large $N_c$ limit counting of the $N_c$-factors reveals mesons
to be stable and non-interacting. Baryons are different. They are
color singlets of $N_c$ quarks, and accordingly their mass scales as
$N_c \Lambda_{QCD}$, where $\Lambda_{QCD}$ is the (only) dimensional
QCD scale factor. Their radius remains proportional to
$1/\Lambda_{QCD}$ and $N_c$ independent. Accordingly the quark
density with the baryon grows beyond bound, and a Hartree
approximation becomes appropriate \cite{Witten}. Finally as the
meson-baryon coupling constants in general are proportional to
$\sqrt{N_c}$, mesons are strongly coupled to baryons in the large
$N_c$ limit.\\

There are two approaches to describe the baryons in a large $N_c$
limit. The first is to employ a constituent quark model description
based on appropriately symmetrized products of $N_c$ quark wave
functions in the Hartree approximation. The other is to construct the
baryons as topologically stable soliton solutions to a chiral
Lagrangian of meson fields, with the general form \cite{Jenkins}
$${\cal L}_{meson}=N_c{\cal L}_p({\phi\over \sqrt{N_c}}),\eqno(1)$$
where ${\cal L}_p$ is polynomial of meson fields and their gradients,
which satisfies the (nontrivial) stability requirements. 
The Skyrme model \cite{Skyrme}, and its
generalizations \cite{Marleau,Jackson} are generic examples of the
latter approach.

\section{Large $N_c$ Operator Algebra}

Among mesons the pions stand out because their role as the Goldstone
bosons of the spontaneously broken approximate chiral symmetry of QCD.
As such their coupling to hadrons has to vanish with 4-momentum. The
pion coupling to fermions is accordingly 
$${\cal L}={1\over f_\pi}A^{\mu a}\partial_\mu \pi^a,\eqno(2)$$
where $f_\pi$ is the pion decay constant $(\sim \sqrt{N_c})$, and
$A^{\mu a}$ is the axial vector of the fermion, which in the case of
a baryon scales as $N_c$. \\

In the large $N_c$ limit, the baryons propagator reduces to the static
propagator, and $\pi$-baryon scattering becomes recoilless.
Consequently the $\pi$-baryon scattering amplitude only involves
bilinear combinations of the space components of the axial current,
from which it proves convenient to separate a factor $N_c$ as
$$A^{ia}=g N_cX^{ia},\eqno(3)$$
where $g$ is a coupling constant introduced for convenience.\\

For the operators $X^{ia}$ the following $N_c$ expansion proves useful
\cite{Manohar}:
$$X^{ia}=X_0^{ia}+{1\over N_c}X_1^{ia}+{1\over
N_c^2}X_2^{ia}+...\eqno(4)$$
The lowest order operator $X_0^{ia}$ and the spin- and isospin
operators then satisfy a contracted SU(4) algebra. This is the algebra
of the Skyrme model, which is made explicit by the identification
$$X_0^{ia}={1\over 2}Tr\{A \sigma^i A^\dagger \tau^a\},\eqno(5)$$
A being the SU(2) rotational collective coordinate used in the spin-isospin
quantization of the Skyrme Hamiltonian \cite{Adkins}.\\

Baryon operators may then described systematically in terms of the
operators $X_n^{ia}$ or alternatively in terms of the SU(4) generators
spin, isospin and spin-isospin $G^{ia}$, where the relation to the
$X_n^{ia}$:s is given by the limiting relation 
$$\lim_{N_c\rightarrow \infty}{G^{ia}\over N_c}=X_0^{ia}.\eqno(6)$$
The connection to the constituent quark model obtains by expression of
the SU(4) generators in terms of corresponding quark operators:
$$J^i=\sum_{l=1}^{N_c}q_l^\dagger{\sigma^i\over 2}q_l,\quad
I^a=\sum_{l=1}^{N_c}q_l^\dagger{\tau^a\over 2} q_l,$$
$$G^{ia}=\sum_{l=1}^{N_c}q_l^\dagger {\sigma^i\over 2}{\tau^a\over
2}q_l.\eqno(7)$$
Thus in the large $N_c$ limit the constituent quark model and the
Skyrme model give equivalent results.

\section{Phenomenological considerations}

The formalism outlined above may be employed to derived systematic
$1/N_c$ expansions for baryon operators as masses, magnetic moments
\cite{Jenkins}. Relations between the available two-body operators
allows expression of the baryon mass operator as
$$M=m_0N_c+m_2{1\over N_c^2}\vec J^2+m_3{1\over N_c^3}\vec
J^4+..,\eqno(8)$$
where $\vec J$ is the spin-operator. To order $1/N_c^2$ this agrees
with the Skyrme model result \cite{Adkins} 
$$M=M_s+{1\over 2\Omega_s}\vec J^2,\eqno(9)$$
where $M_s$ and $\Omega_s$ are the mass and moment of inertia of the
soliton. \\

Returning to the quark model representation, (7) it has been found
that among the two-body operator combinations of $J^i$, $I^a$ and
$G^{ia}$, the only significant combination for the mass operator is
the operator
$$-\sum_{i<j}\vec \sigma_i\cdot \vec \sigma_j\vec \tau_i\cdot \vec
\tau_j,\eqno(10)$$
and its generalization SU(3) flavor \cite{Georgi,Carlson}. This is
consistent as it is this operator and the scalar unit operator, which
have the highest $N_c$-scaling factors \cite{Wirzba}. As this
operator appears in the pion exchange and multipion exchange
interaction between constituent quarks, it suggests that in the region
between the chiral restoration scale $\Lambda_\chi^{-1}\sim 1/4f_\pi$
and the confinement scale $1/\Lambda_{QCD}$ the effective dynamical
description of baryons is in terms of constituent quarks that
interact by exchanging pions \cite{Mano}.\\

The reason for the effectiveness of the operator (8) in organizing the
baryon spectrum in agreement with the empirical one is that it is
the only operator combination that is able to achieve the correct
ordering of the positive and negative parity states in the spectrum.
Any monotonic confining interaction would order the states in baryon
spectrum in shells of alternating parity. Yet the lowest excited
states in the nucleon spectrum are successively the $N(1440),\,{1\over
2}^+$ and the $N(1535),\,{1\over 2}^-$. As these states (and the nucleon)
all have mixed color-spin symmetry $[21]_{CS}$, the color-spin
dependent perturbative gluon exchange model for the hyperfine
interaction cannot reverse their normal ordering and bring them into
agreement with the empirical ordering. On the other hand
the $N(1535),\,{1\over 2}^-$ state has mixed flavor-spin symmetry
$[21]_{FS}$, whereas the nucleon and the $N(1440),\,{1\over 2}^+$ state
have complete flavor-spin symmetry $[3]_{FS}$. Therefore, in
combination with a sufficiently strong orbital matrix element, the
operator (8) is able to reverse the normal ordering, and bring the
spectrum into agreement with experiment \cite{Gloz}. This argument
carries over to all flavor sectors of the baryon spectrum, with
exception of the $\Lambda$-spectrum, where the negative flavor singlet
$\Lambda 1405-\Lambda 1520$ is the lowest multiplet, which however, also
only can be understood with the operator form (8).\\

The interaction (8) may be interpreted as being due to pion
\cite{Gloz} and multipion exchange between the constituent quarks.
Two-pion exchange is required to cancel out the tensor interaction of
one-pion exchange, as the tensor interaction would otherwise imply
substantial spin-orbit splittings in the $P$-shell, in conflict with
experiment \cite{Brown}. With suitable parameter choices
the inclusion of two-pion exchange also allows a
cancellation of the spin-orbit interaction that arises
with a linear scalar confining
interaction.\\

The constituent quark model, with quarks interacting by pion exchange,
is usually referred to as the chiral quark model. The dynamical
interpretation resembles that of the Skyrme model, Hence, and in view
of the discussion in section 2 above, it is no surprise that the
reversal of the normal ordering of the states in the baryon spectrum
also obtains in the Skyrme model. This model has long been known to
imply very low lying vibrational states \cite{Dothan,Mattis}.

\section{Heavy Flavor Baryons and Pentaquarks}

The quark model describes baryons as 3-quark systems. The Skyrme model
describes baryons as formed of $(q\bar q)^n$ - i.e. mesons - in the
field of a soliton, which carries the baryon number. Hyperons are then
best described as bound states of soliton and $K$, $D$ and $B$ mesons
respectively \cite{Callan,Rho}.\\

This provides a unified description of normal hyperons and
"pentaquarks". While a hyperon is described as a bound state of a
soliton and a heavy flavor meson, the corresponding pentaquark is
described as a bound state of soliton and the corersponding heavy
flavor antimeson \cite{Scoc}. Note that the nomenclature is due to the
non-symmetric description by the constituent quark model of the former
as $qq Q$ states and the latter as $qqqq \bar Q$ states, where $q$ and
$Q$ denote light and heavy flavor quarks respectively.\\

Both the chiral quark model, extended to broken $SU(N_F)$
\cite{Gloz2}, and the bound state soliton model \cite{Rho} describe
the ground state heavy flavor hyperons well \cite{NIM}. But only the
latter yields absolute predictions for the pentaquark energies without
further parameters or assumptions, as the difference between hyperon
and pentaquark energies arises solely from the sign of the Wess-Zumino
term in the effective meson-soliton interaction \cite{Callan,Scoc}.\\

The existence or non-existence of pentaquarks is interesting, because
it depends very much on the form of the hyperfine interaction between
quarks. If that interaction has the form of perturbative gluon
exchange, the lowest energy pentaquark contains a strange quark, and
has negative parity \cite{Gignoux,Lipkin}. The bound state soliton
model \cite{Scoc} and the chiral quark model \cite{Stancu} in contrast
predict that the lowest energy - and definitely stable - pentaquark is
non-strange and has positive parity. Only the former has been searched
for experimentally, although with somewhat inconclusive results
\cite{Aitala}. Experimental identification of the latter would be very
informative, and would automatically explain the non-existence of the
$H$-particle, which hasn't been found in spite extensive effort
\cite{Ica}.

\section{Nuclei}

Most of Judah Eisenberg's work on applications of the Skyrme model
dealt with nuclei. This work employed Skyrme's product ansatz for
solitons with baryon number larger that 1 \cite{Skyrme2}:
$$U(r;\, \vec r_1,...\vec r_A)=\Pi_{i=1}^{A} U(\vec r-\vec
r_1)\eqno(9)$$
When this ansatz is inserted into the Lagrangian density of the Skyrme
model, the Lagrangian separates into a sum of single nucleon terms and
interaction terms involving $2,\, 3,\,...A$ nucleons. When integrated
over $\vec r$ the latter yield models for the $2$-,\,
$3$-,\,...$A$-nucleon interactions. For large internucleon separations
these terms have the same form as the corresponding pion-exchange
interactions that obtain with more conventional chiral Lagrangians
\cite{Nyman}.\\

This is illustrated by the derivation of the three-nucleon interaction
based on the product ansatz by Eisenberg and K\"albermann \cite{Judah}.
For large interparticle separations this interaction reduces to the
conventional two-pion exchange three-nucleon interaction, that
involves an intermediate $\Delta_{33}$ resonance in the sharp
resonance approximation \cite{DOR2}. The appearance of the sharp
resonance propagator $-i(m_\Delta-m_N)$, in this expression is a
direct consequence of the large $N_c$ mass formula (9).\\

The product ansatz (9) does not, however, provide a good approximation
to the minimum energy configuration of 8 skyrmions with $B>1$. The
minimum energy solutions have an interesting topology, which hitherto
has not yielded to analytical treatment \cite{Braa,Batt}. This
situation has now changed by the discovery of elegant rational map
approximations to the minimal energy solutions \cite{Conor}.\\

The rational map approximation for the $B=n$ skyrmion takes the form
$$U(\vec r)=e^{i\vec \tau\cdot \vec \pi_nF(r)},\eqno(10)$$
where the "chiral angle" $F(r)$ depends only on the distance to the
center, and $\vec \pi_n$ is defined as
$$\vec \pi_n={1\over 1+|R_n(z)|^2}\{2Re[R_n(z)],\quad
2Im[R_n(z)],\quad 1-|R_n(z)|^2\}.\eqno(11)$$
Here $R_n(z)$ is the rational map for baryon number $n$, and $z$ is
defined as $z=tan v/2\, e^{i\rho}$. For $n=1$, $R(z)=z$, and (10)
reduces to Skyrme's hedgehog solution. For larger $n$ $R(z)$ are simple
rational functions of $z$, the simplest case being $n=2$ for which
$R_2(z)=z^2$.\\

The rational maps open the door to more realistic applications to
nuclei based on the Skyrme model, than what hitherto has been
possible. As an example the question of the existence of bound states
between $\eta$-mesons and nuclei may be addressed and shown to be
likely \cite{Scoc3}. What is still wanting, however, are functional
forms that connect the rational maps approximations (10) to the product
ansatz (9), which is appropriate for large separations.

\section{Judah}

Juhah would have enjoyed the elegance of the rational maps. Of his
style in expression, here he is in a letter, dated November 12,
1973, in Charlottesville: ``Stanley Hanna and I were very much
hoping that you might be able so see your way clear to give a
talk at the [1974 photonuclear Gordon] conference. In particular,
we wondered if you would be willing to review ... . Of course,
it would be appropriate to slant your discussion towards
aspects of this problem, which would be of particular
relevance for the community of photonuclear physicists''.
Obviously I accepted. That photonuclear conference, which
Judah chaired, will be long remembered by the 
participants, as one of the (unscheduled) evening talks
was the televised resignation speech of R. M. Nixon.

%
%

\end{document}